# Network Traffic Management

Dr. Namdeo V. Kalyankar

**Abstract**—The purposes of this paper have to discuss issues related to Network Traffic Management. A relatively new category of network management is fast becoming a necessity in converged business Networks. Mid-sized and large organizations are finding they must control network traffic behavior to assure that their strategic applications always get the resources they need to perform optimally. Controlling network traffic requires limiting bandwidth to certain applications, guaranteeing minimum bandwidth to others, and marking traffic with high or low priorities. This exercise is called Network Traffic Management.

—————————— ◆ ——————————

## 1 INTRODUCTION

A Computer network is a data communications system which interconnects computer systems at various different sites. A network may be composed of any combination of LANs, or WANs.

Network traffic can be defined in a number of ways. But in the simplest manner we can define it as the density of data present in any Network. In any computer Network, there are a lot of communication devices trying to access resources and at the same time getting requests to carry out some work for some other device [1]. Also at the same time certain types of communication devices may be busy to respond to the request being made to them. So there is lot of information exchange in the Network in form of request, response and control data. This data is basically in the form of a huge number of packets floating around in the Network. This huge amount of data acts as a load on the Network, which results in slowing down the operations of other communication devices. Due to this there is a lot of delay in communication activities. This ultimately results in congestion of the Network. This is the description of Network Traffic in its simplest form. In other words we can say that Network traffic is the load on the communication devices and the system.

This traffic on the network has now resulted in mid-sized and large organizations realizing that they must control network traffic behavior to ensure that their strategic applications always get the resources they need to perform optimally [2]. Controlling network traffic requires limiting bandwidth to certain applications, guaranteeing minimum bandwidth to others, and marking traffic with high or low priorities. This exercise is called traffic management.

## 2 GENERAL PROCESSES FOR TRAFFIC MANAGEMENT

Traffic Management consists of the amalgamation of a

————————————
- *Dr. Namdeo N. Kalyankar is Principal working with Yeshwant Mahavidyalaya, Nanded, India*

number of activities as shown below:

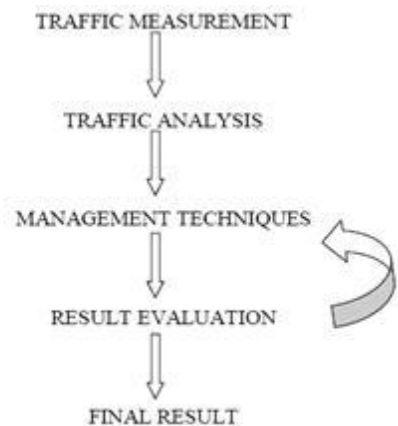

Fig. 1 General Process for Traffic Management

## 3 TECHNIQUE FOR MEASURE NETWORK TRAFFIC

One of the easiest ways to comprehend Network Traffic is to consider an analogy with the road traffic. Consider that there is an emergency and someone has fallen sick and has to be rushed to the hospital. But when the ambulance tries to make its way through the roads of the city, it finds the roads totally blocked with cars n busses. The solution to this situation would be for a traffic policeman to step in and manage the traffic. He will first gauge the traffic, and then prioritize the traffic [3]. The ambulance will get the highest priority and the road will be made empty for the ambulance to pass. Similar is the case with Network Traffic. When you send a request on the network, it is possible that due to some problem or other requests you have to wait for some time. If over a period of time a number of packets queue up and wait then it results in traffic. Once traffic is created, you must wait till it is over, which can be for any length of time, depending on the situation. So, there has to be some way to deal with this situation. The solution for this is Network Traffic Management and this process starts first with measuring the traffic on the network [4].

### 3.1 Reasons to measure network traffic:

The following are the resons for wich we have measure the network traffic.
a) Service monitoring - making sure things keep working.
b) Network planning - deciding when more capacity is needed.
c) Cost recovery - session times and traffic volumes can provide billing data.
d) Research - an improved understanding of what's happening should allow us to improve network performance.

### 3.2 Internet traffic

The basic performance metrics of Internet traffic can be listed as:-
- Packet loss
- Delay
- Throughput
- Availability

### 3.3 Drivers for measurement

There are number of other drivers strongly deals with requirement of measurement are –
- Pricing
- Service level agreements
- New services
- Applications

## 4 NETWORK TRAFFIC MEASURE

Usually, traffic management is deployed at the WAN edge of an enterprise site. This is where the high-speed LAN meets the lower-speed WAN access link. The LAN-WAN juncture is also where both Internet and intranet traffic enter and exit the enterprise [5]. So it is the ideal place to "tame" traffic and to mitigate the impact of non-critical and even suspicious traffic picked up on the Internet. Limiting or blocking the network resources available to frivolous or undesirable traffic boosts the performance of enterprise resource planning (ERP), customer relationship management (CRM), and other strategic, business-critical applications.

In addition to monitoring traffic at the network edge, there are pure performance issues to consider. The WAN access network is usually slower than the LAN, generally for budgetary reasons. Also Businesses pay recurring monthly fees for WAN services, while LAN bandwidth is free (after the initial equipment investments have been made). With high-speed LAN traffic slowing down at the lower-speed access circuit, the LAN-WAN edge is where congestion is most likely to occur. Another important factor to consider here is that most applications have been developed to run on LANs. Now, local networks are generally free from congestion and fall under the total control of an internal IT department [6]. These LAN-optimized applications behave differently in the WAN environment. Not only is the WAN access link slower, but WAN service also can fall under the management purview of multiple network providers. Managing traffic in this network segment aids distributed organizations that depend on the WAN to serve remote users with centralized resources. Doing so is a reasonably simple matter. In most cases, a network administrator uses a GUI to set parameters for some business-critical policies in plain English. The administrator then pushes a button to propagate those policies to the various network segments where they should be enforced.

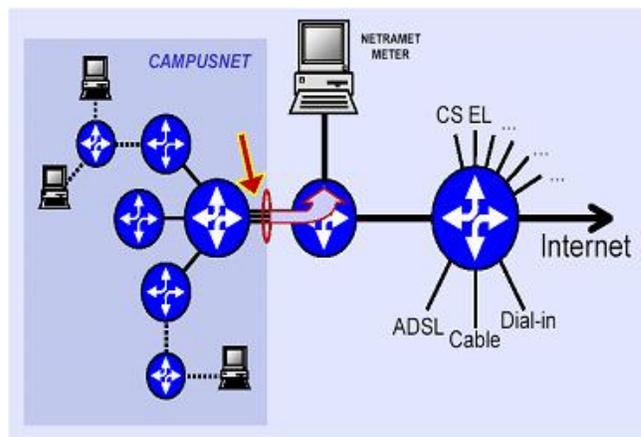

Fig. 2 Traffic Management at the WAN edge of an Enterprise

## 5 APPROACHES FOR TRAFFIC MEASUREMENT

There are some approaches for traffic measurement as follows

### 5.1 Active Measurement of Traffic

As name indicates, in this measurement approach users or providers are directly related to the activities to the measurement[5]. There are number of different ways to carry out this measurement like
i) Injection of probes into network by users and providers
ii) Ping and Trace out the Path connectivity and Round-trip delay
iii) User-application performance as seen from hosts
like Loss, Delay and Throughput
iv) Distribute on measurement servers makes the Probes are spread across mesh of paths through network to check scalability and growth of probe traffic

### 5.2 Passive Measurement of Traffic

In this approach user is indirectly deal with system using some hardware or software tools. Basically some historical data is used to find the current traffic measurement [7]. The currently used techniques for this type of measurement are as follows
i) Packet monitors: This can be achieved by recording packet headers on link. It requires unique detail of protocol and architecture studies
ii) Router / Switch traffic statistics: Analyzing router or switch, the intelligent devices installed at network, can provide network internal behavior. Using these devices we can get information about Packet drops, Counts and Flow statistics

iii) Server and router logs: These records or logs can perform well work in measuring. They provide summaries of dial session, routing updates or web-server log.

## 6 MEASURING TOOLS

There are many tools available for measurement of traffic [7]. They are listed according their categories. The Local Systems which includes NETSTAT, TCPDUMP, ETHREAL and NTOP. The Remote (END) System which having MIB, IF-MIB, SNMP and MRTG . The Routers are also having NETFLOW (CISCO) and LFAP (ENTERASYS). Lastly the SNIFFERS having RMON, RMON2 and NETRAMET

## 7 TRAFFIC ANALYSIS

After consecutive monitoring over a number of years, LAN and WAN traffic have been seen to follow different patterns.

### 7.1 Lan Traffic:

Traffic on a LAN has shown to be self similar in nature [7]. Those means if I measure the traffic over a period of one hour and plot it, it will be similar to the graph for the traffic plotted over a period of one day. In the same manner the day graph will be similar to the traffic graph plotted over a week and the week graph for that of a month. The patter of the variation of the traffic repeats itself over regular intervals.

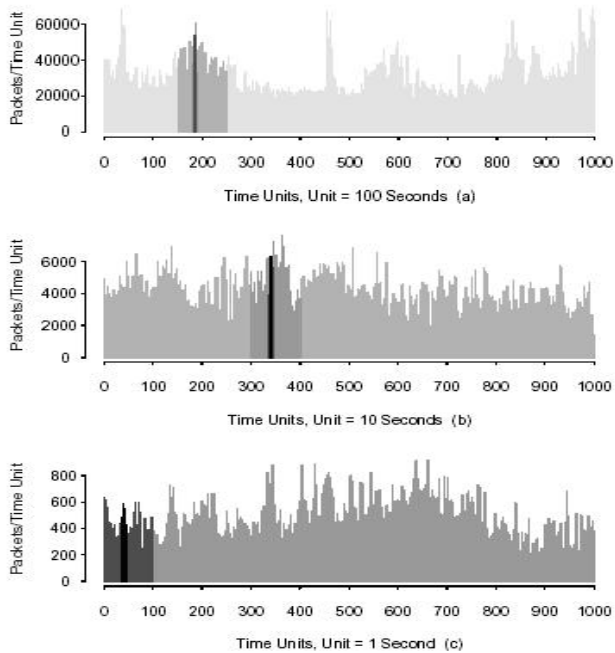

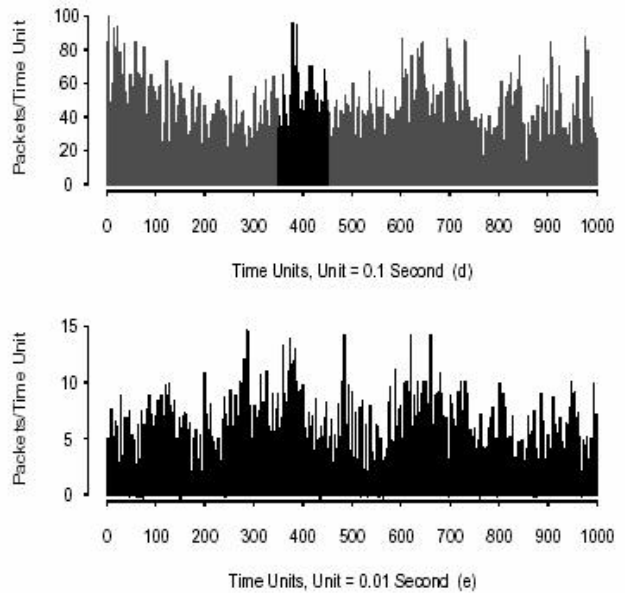

Fig. 3 (a)—(e). Pictorial "proof" of self-similarity: Ethernet traffic (packets per time unit for the August '89 trace) on 5 different time scales. (Different gray levels are used to identify the same segments of traffic on the different time scales.)

### 7.2 WAN Traffic:

Traffic on the WAN has been found to vary as per the following models [8].
*Random Traffic*: The traffic here seems to follow no fixed pattern.

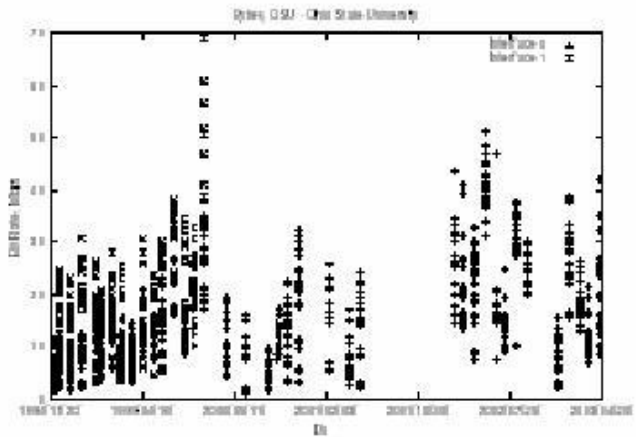

Fig. 4 Fixed Pattern

*Poisson's Model*: Traffic Nature in Internet has been identified to confirm to the Poisson's Model. This model gives us a rough idea of the characteristics of Internet Traffic. The model estimates the probability of the number of packets that should be present on the network after a given time if the average arrival rate of the packets is specified.
*Bursty Traffic*: This model states that, the average traffic over the network stays roughly constant, except for the sudden bursts (long and short).

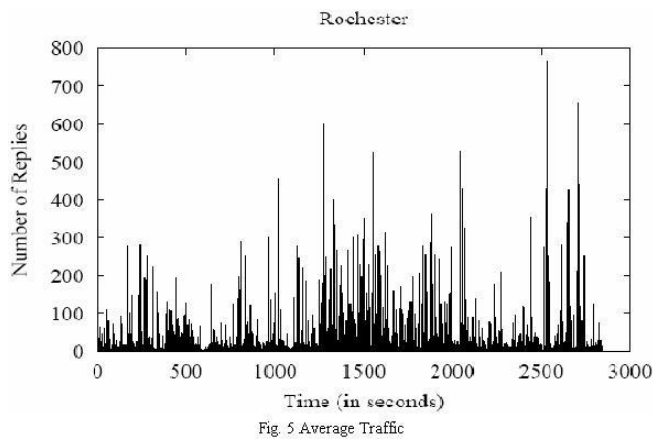

Fig. 5 Average Traffic

## 8 TRAFFIC MANAGEMENT

A look at the figure below will make the comprehension of network traffic before and after it is managed more clear [8]. The figure is a depiction of the transmission media whilst it is carrying the unmanaged traffic. As we can see normal applications such as maybe video, audio downloads etc are taking up the major portion of the available band width. Mission critical applications are left with only about 40 % bandwidth which means that there may be a lot of delay in the transmission data or processing of transactions. This is where the role of traffic management comes in.

The user can take a decision as to how much amount of bandwidth he wants to keep exclusively for mission critical applications, and then the rest can be used for other normal applications [9]. In the second figure we can see that the traffic has been managed in such a way that maximum bandwidth (nearly 70%) has been reserved for mission critical applications. 5% of bandwidth is unused which can also be used by these applications in case of a surge in traffic. Normal applications are left with only about 25% of the band width.

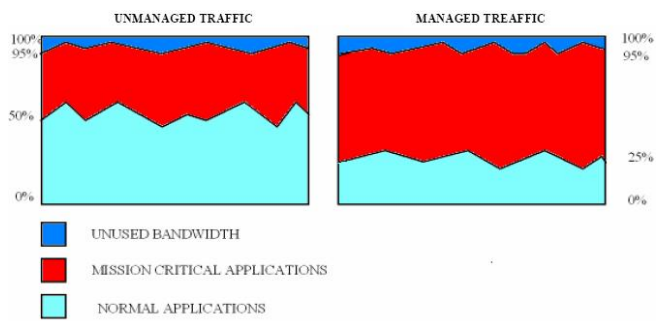

Fig. 6 Traffic Management

## 9 CONCLUSION

To conclude we would like to reemphasize that, in today's changing scenario, where the conventional way of doing things no longer holds good, organizations are fast realizing that in order that they stay in step with others in the race, they must embrace this concept of Network Management. Also the manner in which both the size of networks and the data which rides on them is increasing by the day, it has become imperative to monitor the kind of traffic flowing, priorities it and then manage the traffic accordingly.

### FIGURES

Fig. 1 General Process for Traffic Management
Fig. 2 Traffic Management at the WAN edge of an Enterprise
Fig. 3 Prictorial proof of self-similarity
Fig. 4 Fixed Pattern
Fig. 5 Average Traffic
Fig. 6 Traffic Management

### AUTHORS

**Namdeo V. Kalyankar:** Completed M.Sc. Physics from B.A.M. University , Aurangabad. in 1980. in 1980 he joined as Lecturer in Department of Physics in yeshwant College,Nanded. In 1984 he completed his DHE. He Completed his Ph.D. from B.A.M.University in 1995. From 2003 he is working as Principal since 2003 to till date in Yeshwant college Nanded. He is also Research Guide for Computer Studies in S.R.T.M. University , Nanded. He is also worked on various bodies in S.R.T.M. University Nanded. He also published research papers in various international/ national journals. He is peer team member of NAAC (National Assessment and Accreditation Council)(India). He published a book entitled " DBMS Concept and programming in Foxpro". He also got "Best Principal" award from S.R.T.M. University, Nanded(India) in 2009. He is life member of Indian National Congress , Kolkata (India). He is also honored with "Fellowship of Linnean Society of London (F.L.S.)" on 11[th] Nov. 2009